\documentclass[prl,onecolumn.showpacs,preprintnumbers,amsmath,amssymb]{revtex4}
\usepackage{graphicx}
\usepackage{dcolumn}
\usepackage{bm}

\begin{document}

\title{Self-cooling of a micro-mirror by radiation pressure}

\author{
S. Gigan$^{1,2}$, H. R. B\"ohm$^{1,2}$, M.
Paternostro$^{2,\dagger}$, F. Blaser$^{1,2}$, G. Langer$^{3}$, J.
B. Hertzberg$^{4,5}$, K. Schwab$^{4,\ddag}$, D. B\"auerle$^{3}$, M.
Aspelmeyer$^{1,2}$\footnote{Corresponding author:
markus.aspelmeyer@quantum.at}, A. Zeilinger$^{1,2}$}
\affiliation{$^1$~Physics Faculty, Institute for Experimental Physics, University of Vienna, Boltzmanngasse 5, A-1909 Vienna, Austria\\
$^2$~Institute for Quantum Optics and Quantum Information (IQOQI), Austrian Academy of Sciences, Boltzmanngasse 3, A-1090 Vienna, Austria\\
$^3$~Institute for Applied Physics, Johannes-Kepler-University Linz, Altenbergerstr. 69, A-4040 Linz, Austria\\
$^4$~Laboratory for Physical Sciences, University of Maryland, College Park, MD 20740 USA\\
$^5$~ Department of Physics, University of Maryland, College Park, MD 20740, USA\\
$^{\dagger}$~permanent address: School of Mathematics and
Physics, Queen's University Belfast, UK\\
$^{\ddag}$~present address: Cornell University, USA
}

\date{\today}

\begin{abstract}
We demonstrate passive feedback cooling of a mechanical resonator
based on radiation pressure forces and assisted by photothermal forces
in a high-finesse optical cavity. The resonator is a free-standing
high-reflectance micro-mirror (of mass $m\approx 400~ng$ and
mechanical quality factor $Q\approx 10^4$) that is used as
back-mirror in a detuned Fabry-Perot cavity of optical finesse
$F\approx500$. We observe an increased damping in the dynamics of
the mechanical oscillator by a factor of 30 and a corresponding
cooling of the oscillator modes below 10~K starting from room
temperature. This effect is an important ingredient for recently
proposed schemes to prepare quantum entanglement of macroscopic
mechanical oscillators.
\end{abstract}

\maketitle

\section{Introduction}

Cooling of mechanical resonators is currently a hot topic in many
fields of physics including ultra-high precision measurements
\cite{LaHaye2004}, detection of gravitational waves
\cite{Braginsky2002,LIGO} and the study of the transition between
classical and quantum behavior of a mechanical system
\cite{Schwab2005,Leggett2002,Bouwmeester2003}. Here, we report the
first observation of self-cooling of a micro-mirror by radiation
pressure inside a high-finesse optical cavity. In essence, changes
in intensity in a detuned cavity, as caused by the thermal vibration
of the mirror, provide the mechanism for entropy flow from the
mirror's oscillatory motion to the low-entropy cavity field
\cite{Braginsky2002}. The crucial coupling between radiation and
mechanical motion was made possible by producing free-standing
micro-mirrors of low mass (m $\approx$ 400 ng), high reflectance
($>$99,6\%)and high mechanical quality ($Q\approx10^4$). We observe
cooling of the mechanical oscillator by a factor of more than 30,
i.e. from room temperature to below 10~K. In addition to purely
photothermal effects \cite{Metzger2004} we identify
radiation-pressure as a relevant mechanism participating to the
cooling. In contrast to earlier experiments, our technique does not
need any active feedback \cite{Mertz1993,Cohadon1996,Bushev2006}.
Our results suggest that it should be possible to reach very low
temperatures. We expect that improvements of our method will allow
for pure radiation pressure cooling, with cooling ratios beyond
1.000, and thus possibly enable cooling all the way down to the
quantum mechanical ground state of the micro-mirror.

Radiation pressure forces inside optical cavities are known to
pose an ultimate limit on the sensitivity of interferometric
measurements \cite{Caves1980,Braginsky2001}. However, less known,
radiation pressure can also be used for the opposite, namely to
counteract the dynamics of a cavity mirror via dynamical back
action \cite{Braginsky2002,Metzger2004,Braginsky1970}.  In a
recent experiment Metzger and Karrai \cite{Metzger2004} presented
a passive cooling mechanism for a micro-mechanical oscillator
based on bolometric back action. Even though this scheme has
intrinsically limited cooling capability since it ultimately
relies on heating by absorption, it may allow for a
quantitatively significant reduction of the oscillator's thermal
motion. A more powerful scheme is provided by the use of
radiation pressure as a feedback force \cite{Braginsky2002}. In
this case, optical absorption does not impose a fundamental
limit. The difficulty in utilizing radiation pressure for this
cooling purpose is that it requires stable control of the
detuning of a high-finesse cavity, strong optomechanical coupling
and a low mass of at least one cavity mirror, hence nano- or
micro-mechanical systems of high optical and mechanical quality
(characterized by the cavity finesse F and the mechanical quality
factor Q). Although cavity-induced radiation-pressure effects have
already been used to modify elastic properties of mirrors
\cite{Metzger2004,Tucker2002,Vogel2003,Sheard2004} and to enforce
mechanical instabilities
\cite{Tucker2002,Dorsel1983,Rokhsari2005,Kippenberg2005,Kippenberg2005b},
none of the previous experiments was able to combine these strict
requirements. We have overcome this limitation by developing a
method to produce free-standing micro-mirrors of low mass
($Q\approx400ng$) high reflectance ($>$99.6\%) and high mechanical
quality ($Q\approx10^4$). Using such micro-mirrors in a detuned
optical cavity allows us to observe for the first time
self-cooling in a regime where, although photothermal effects are
still present, radiation pressure significantly participates  in
the self-cooling process.

\section{Idea of radiation-pressure cooling}

Radiation pressure forces in an optical cavity arise due to the
momentum transfer of photons reflected from the mirror surface. For
certain cavity detuning, i.e. if the cavity angular frequency
$\omega_c$ is off resonance with the frequency $\omega_l$ of the
pump laser, the radiation pressure is highly sensitive to small
displacements of the cavity mirror. This is a consequence of the
fact that the energy stored in a cavity field varies strongly with
detuning. As a consequence, the dynamics of an oscillating mirror
inside a detuned cavity is modified by a mechanical rigidity that
depends on the detuning. For a high-finesse cavity, the
radiation-pressure induced back action can act on the mirror motion
in a way to induce low noise damping. This is the general concept of
dynamical back action which has first been introduced by Braginsky
\cite{Braginsky1970}. A simple classical description of the dynamics
of the mirror shows that both the resonance frequency $\omega_M$ and
the natural damping rate $\gamma$ of the mirror motion are modified
by radiation pressure to $\omega_{\text{eff}}$ and
$\gamma_{\text{eff}}$, respectively
\cite{Braginsky2002,Metzger2004}. In particular, within the
classical framework, the modified damping rate follows
\begin{equation}
\label{damping}
\gamma_{\text{eff}}=\gamma+\frac{\beta(\Delta)}{2m}\frac{2 \kappa}{(2 \kappa)^2+\omega_M^2}
\end{equation}
with the cavity decay rate $\kappa={\pi{c}}/{2 F L}$, the cavity
finesse $F$, the cavity length $L$ and the vacuum speed of light
$c$. Optimum damping is achieved when $1/\kappa$ is of the order
of $\omega_M$, which for $\omega_M$  in the MHz range requires a
high finesse cavity. Equation~(\ref{damping}) depends on
$\beta(\Delta)$, the spatial gradient of the radiation force
evaluated at a (spatial) detuning
$\Delta_x=L{\Delta}/{\omega_l}$. Here, $\Delta$ is the effective
detuning between cavity and laser frequency, including the effect
of radiation pressure \cite{Giovanetti2001}. The contribution
$\beta$, induced by radiation pressure, can be positive or
negative depending on the sign of $\Delta$. It is straightforward
to show that $\beta(\Delta)$ is negative for $\Delta<0$,
corresponding to $\gamma_{\text{eff}}<\gamma$. In this regime,
the system can enter into instability. The focus of this work is
the investigation of the opposite regime  ($\beta(\Delta)>0$) in
which $\gamma_{\text{eff}}>\gamma$. This low-noise damping
results in a reduction of the mirror temperature and hence
self-cooling is achieved. The previous self-cooling experiments
based on bolometric forces \cite{Metzger2004} were operated in
the regime of negative detuning where radiation pressure
counteracts the cooling.

To observe the self-cooling effect a read-out scheme of the mirror
motion is required. To do that it turns out that it is sufficient to
measure the statistical properties of the optical field that leaks
out of the cavity. In a way, the output cavity field represents a
"blank paper" on which the dynamics of the mirror can be written. It
is possible to briefly sketch the main idea of our self-cooling
read-out process by exploiting a simple (but for our purposes
sufficient) semiclassical picture. A full quantum mechanical
framework, which generalizes the classical picture for self cooling
proposed so far in the literature \cite{Braginsky2002,Bushev2006},
is presented elsewhere \cite{Paternostro2006}. Not only is this
(more general) approach in agreement with the classical picture
taken into account by Eq. \ref{damping} but it also paves the way
toward the rigorous study of the limitations imposed to self cooling
by the influences of quantum noise \cite{Paternostro2006}. The total
energy of a cavity consisting of a fixed mirror and a movable mirror
driven by an input laser field of power P is given by
\cite{Giovanetti2001,Braunstein2003}
\begin{equation}
\label{energy}
E=\hbar(\omega_c-\omega_l)(X^2+Y^2)-\hbar\frac{\omega_c}{2 L}(X^2+Y^2-1){q}+\frac{1}{2}\left(\frac{{p}^2}{m}+m\omega^{2}_{M}{q}^2\right)+\sqrt{2} \hbar{\cal
E}Y,
\end{equation}
where $X$ and $Y$ are the quadratures of the cavity field, $p$ and
$q$ are the momentum and position quadratures of the oscillating
mirror, and  ${\cal E}=\sqrt{2\kappa{P}/\hbar\omega_l}$ is the
coupling rate between the cavity and the input laser field. If the
time-scale set by the cavity decay rate is the shortest in the
dynamics of the system, i.e. $\kappa\gg\omega_M$, the cavity field
follows the mirror motion adiabatically. As a consequence, the
fluctuations $\delta{Y}_{\text{out}}$ of the field leaking out of
the cavity are directly related to the fluctuations of the mirror's
position quadrature as  $\delta{Y}(t)={\cal A}(\Delta,\kappa,{\cal
E})\delta{q}(t)$\cite{Paternostro2006,Jacobs1999}, where we have
neglected any noise in the system. For the parameter regime of our
experiment the signal-to-noise ratio of the contribution given by
the mirror's spectrum is as large as $10^{7}$. The dynamics of the
output field quadrature is thus entirely determined by the mirror
motion via the function ${\cal}(\Delta,\kappa,{\cal E})$. Therefore,
a phase-sensitive measurement of the output field quadrature
$\delta{Y}_{\text{out}}$ is capable of "monitoring" the full mirror
dynamics. It is particularly interesting to measure the power
spectrum, since
$S_{Y_{\text{out}}}=\int{d}t'e^{i\omega{t}'}\langle{\delta{Y}_{\text{out}}(t)\delta{Y}_{\text{out}}(t+t')}\rangle={\cal
T}(\Delta)S_q$, where $S_q$ is the spectrum of the mirror motion. In
other words, the quadrature power spectrum of the mirror motion
$S_q$ and of the output cavity field $S_{Y_{\text{out}}}$ are
directly related via a transfer function ${\cal T}(\Delta)$. This
correspondence is at the basis of our readout-scheme. Note that the
full transfer function has to take into account the sensitivity of
the specific detection scheme used. This detection strategy allows
us to infer the effective temperature of the mirror Brownian motion
through the study of its displacement power-spectrum, i.e. its
frequency-dependent mean square displacement. The power spectrum
follows a Lorentzian distribution centered around
$\widetilde{\omega_0} =
\sqrt{\omega_{\text{eff}}^2-2\gamma_{\text{eff}}^2}$ with a full
width at half maximum (FWHM) $w_{\text{FWHM}}\approx
2\gamma_{\text{eff}}$ (for $\omega_0^2\gg\gamma_{\text{eff}}^2$),
thus proportional to the introduced damping. The area of the power
spectrum,  $\langle
x^2\rangle=\int_{-\infty}^{+\infty}d\omega{S_q}{}$, is proportional
to the mean energy $\langle E\rangle$ of the vibrational mode and
hence, via the equipartition law, to the effective temperature of
the mirror, since $\langle E\rangle=m\omega_M^2\langle
x^2\rangle=k_BT_{\text{eff}}$ . The relative change in area
underneath the power spectrum is therefore a direct measure for the
change in effective temperature.

\section{Experimental Results and Discussion}

\begin{figure}[h!tbp]
\centerline{\includegraphics[width=0.5\textwidth]{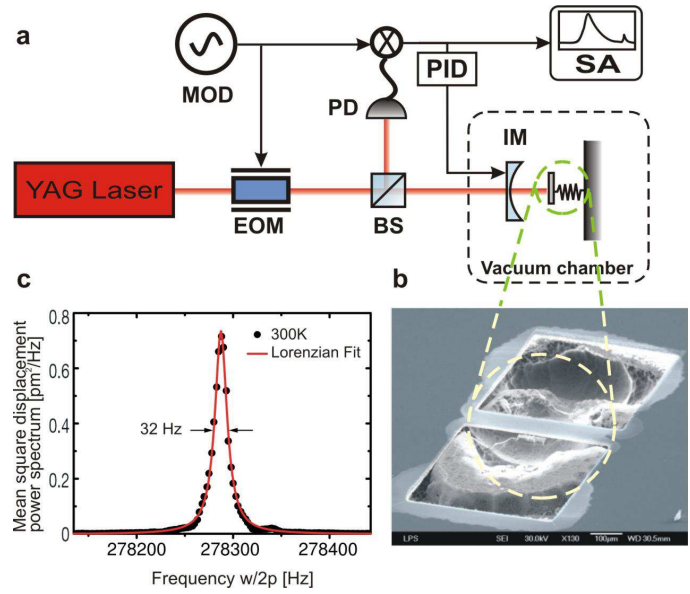}}
\caption{  Sketch of the experimental setup. (a) A cavity is
built between the cantilever and a regular concave mirror of
25~mm focal length and 99.3\% reflectivity. The cavity length was
slightly shorter than 25~mm such as to obtain a waist of
approximately 20~$\mu$m at the location of the surface of the
cantilever. In this configuration we measured a cavity finesse of
500. To minimize damping of the mechanical mode due to gas
friction, the cavity is placed in a vacuum chamber which is kept
at $10^{-5}$ mbar. The cavity is pumped with a Nd:YAG laser at
1064~nm. The beam is phase modulated at 19~MHz (MOD) by a
resonant electro-optic modulator (EOM) before it is injected into
the cavity via the input mirror (IM). The beam reflected from the
cavity is sent via a beam splitter (BS) onto a high-speed PIN
photodiode (PD). After amplification of the photocurrent, its AC
part is demodulated with the initial modulation frequency to
obtain the PDH error signal. This error signal is then used to
feed a low-frequency control loop (PID) to stabilize the cavity
length via a piezo actuator. In addition, the error signal is fed
to a spectrum analyzer (SA) to record the dynamics of the
mechanical mode. (b) The cantilever is a doubly clamped free
standing Bragg mirror (520~$\mu$m long, 120~$\mu$m wide and
2.4~$\mu$m thick) that has been fabricated by using UV
excimer-laser ablation in combination with a dry-etching process
\cite{Baeuerle2000}. The reflectivity of the Bragg mirror is
99.6\% at 1064 nm. (c) Power spectrum of the micro-mirror. We have
isolated a mechanical mode at 280 kHz with a natural width of
32~Hz, corresponding to $Q\approx9000$. All measurements
presented in this work have been made on this mode.
}\label{ExpSetup}
\end{figure}

The system under investigation is a doubly clamped cantilever used
as the end mirror of a linear optical cavity driven by an ultra
stable Nd:YAG laser (see figure \ref{ExpSetup}). The input mirror of
the cavity is attached to a piezo electric transducer which is fed
by a control loop allowing us to lock the precise length of the
cavity either at resonance or detuned (off resonance) with respect
to the laser frequency. The error-signal input to the control loop
is obtained via the Pound-Drever-Hall (PDH) technique
\cite{Black2001}. It has been shown \cite{Jacobs1999} that the PDH
error signal is proportional to the phase quadrature of the output
field ${Y}_{\text{out}}$ and hence to the mirror motion (see above).
An intuitive way to view it is that the error signal is proportional
to the variation of the cavity length. Above the cut-off frequency
of our control loop, the fluctuations in the error signal are
therefore directly related to the thermal noise of the cantilever
(the input mirror is assumed to be fixed).

\begin{figure}[htbp]
\centerline{\includegraphics[width=0.5\textwidth]{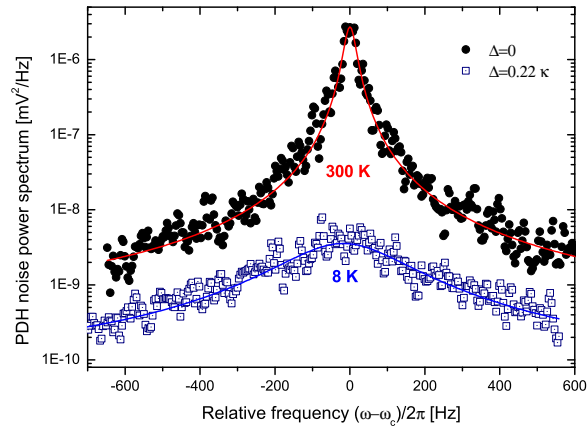}}
\caption{Power spectrum of the mechanical mode at two different
relative detuning levels $\Delta$ of the cavity for an input power
of 2~mW. The data is obtained from the PDH power spectrum, which
is directly proportional to the displacement power spectrum of the
micro-mirror. Experimental points are taken with the spectrum
analyzer, averaged over 30 consecutive measurement runs. Solid
lines are Lorentzian fits to the data. The areas obtained from
the fit correspond to temperatures of 300 K and 8 K,
respectively.} \label{single_power_spectrum_log}
\end{figure}

\begin{figure}[htbp]
\centerline{\includegraphics[width=0.5\textwidth]{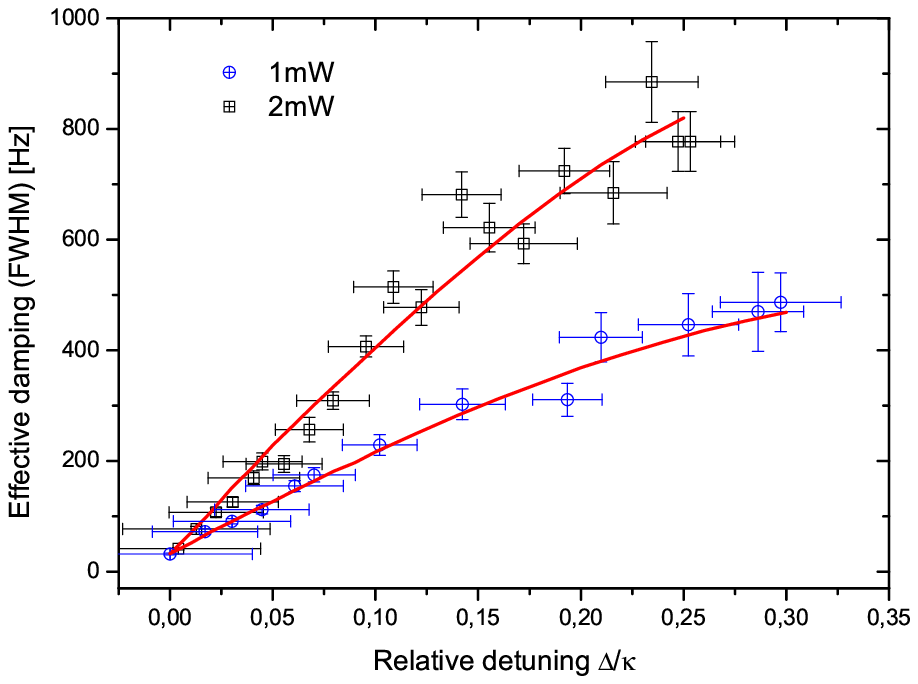}}
\caption{Radiation-pressure induced damping of mirror dynamics. We
show the measured width of the mechanical mode at 278 kHz at
different detuning levels of the cavity and for input laser powers
of 1 mW and 2 mW. The data is obtained directly from Lorentzian
fits on the measured power spectra of the PDH error signal. Error
bars represent absolute errors based on experimental uncertainty.
Solid lines represent theoretical predictions of purely
radiation-pressure effects for $F= 500$, $Q = 9000$, an effective
mass of 9 ng and input powers of 1 mW and 2 mW, respectively. The
inferred effective mass of 22±4 ng indicates the presence of an
additional damping force of photothermal nature (see text). }
\label{single_power_width_log}
\end{figure}

\begin{figure}[htbp]
\centerline{\includegraphics[width=0.5\textwidth]{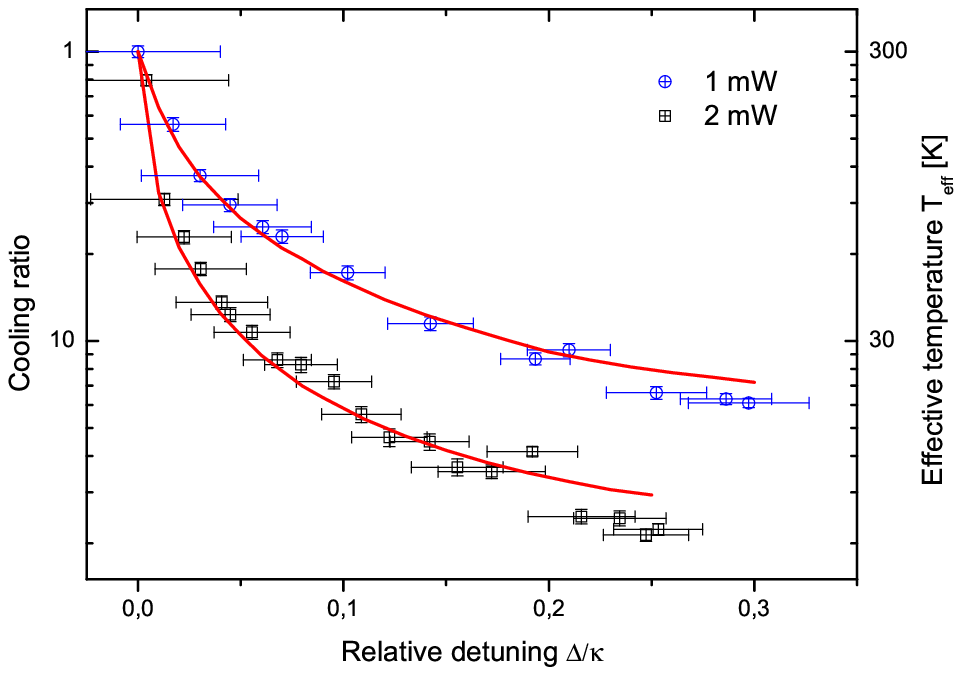}}
\caption{Self-cooling of the mechanical resonator. We show the
cooling ratio on the mechanical mode as a function of detuning and
for input laser powers 1~mW and 2~mW. The data is obtained as
normalized area of the measured PDH power spectrum, compensated
for the detuning dependent sensitivity of the PDH cavity
response. Error bars represent absolute errors based on
experimental uncertainty.  The self-cooling effect increases for
increasing laser power and detuning, in agreement with the
theoretical predictions (solid lines). The right axis shows the
inferred effective temperature of the mechanical oscillator.
Radiation pressure contributes between 30\% and 50\% to the
overall cooling, which is assisted by photothermal effects.
\label{Temperature}}
\end{figure}

We measured the PDH power spectrum for different input powers and
cavity detunings. The detuning was achieved by adding an offset to
the error signal. With this method, the mechanical damping can be
directly measured by determining the FWHM of the resonance peak of
the observed mechanical mode. To obtain the effective temperature of
the mode one has to calculate the area underneath the resonance peak
and to account for the sensitivity of the error signal. This is done
by normalizing the measured mirror amplitudes by the gradient of the
PDH signal. The results are summarized in Figures
\ref{single_power_spectrum_log}, \ref{single_power_width_log} and
\ref{Temperature}.

Figure \ref{single_power_spectrum_log} shows the noise spectrum of
the oscillator for two different detunings at 2~mW input laser
power. The width of the peak increases and the area of the peak
decreases, indicative for both overdamping and cooling of the
mechanical mode. This behavior is in full agreement with the
theoretical model presented above. We investigate the specific
variation of both mechanical damping and of self-cooling with
detuning for different input laser powers of 1~mW and 2~mW,
respectively (Figures \ref{single_power_width_log} and
\ref{Temperature}). Figure \ref{single_power_width_log} shows the
change in width of the mechanical mode. For positive detunings,
the peak is broadened from a natural width of 32~Hz to well above
800~Hz corresponding to an extra damping of the mode. At large
detuning values the stability of the locking limits the precision
of the measurements. For negative detuning (not shown), we
observed a narrowing of the peak, associated with an
amplification of the mirror motion (i.e. "negative" damping),
which rapidly leads to a self-oscillation region. In Figure
\ref{Temperature} the same data set is used to obtain the
corresponding cooling ratio from the relative change in area of
the power spectrum, since the total peak area is a measure of
temperature. As expected, the increase in damping is accompanied
by a cooling of the mechanical mode. At large detuning, the
cooling-effect is slightly enhanced compared to our simple model,
which can be due to the reduced the contribution of thermal
background of other oscillator modes. The best experimental
cooling ratio in our detuning range is above 30. Since our
experiment is performed at room temperature, this corresponds to
a cooling of the mode from 300~K to below 10~K (Fig.
\ref{single_power_spectrum_log}).

We explicitly compare the experimental results for positive
detuning with the theoretical predictions obtained if the effect
is due only to radiation pressure. To do that, we have
independently evaluated the effective mass participating to the
dynamics of the system, which leaves no free parameter for the
evaluation of radiation-pressure forces and hence allows a full
quantitative treatment. The effective mass can be much smaller
than the total mass of the cantilever
\cite{Kippenberg2005,Pinard1999}. ].  For our mirror, an
independent assessment both via spatial tomography of the
vibrational mode and via a calibrated reference results in a
value of 22 ± 4 ng at the probing point (see Appendix). This
results in a theoretically expected cooling less strong than the
experimentally observed one. To get a clear, immediate figure of
the "strength" of the radiation pressure effect required to
replicate the experimental data we assume a fixed effective mass
and allow for variation of the input power. . We find that, for
an effective mass of 18 (26) ng, a power 2.2 (3.3) times larger
than the nominal value used in the experiment is required in
order to match the theoretical predictions with both the observed
damping and cooling  In other words, radiation pressure accounts
for at least 30\% of the observed cooling but may be as strong as
50\%, i.e. cooling by a factor between 8 and 12.  We attribute the
additional cooling in our setup to the presence of photothermal
effects. Similar to the bolometric forces reported in
\cite{Metzger2004}, differential heating of the outer layers of
the dielectric Bragg mirror can result in time-delayed changes of
the cavity length eventually introducing a retarded force that
can contribute to the self-cooling mechanism. In a thin-layered
medium the delayed force induced by photothermal effects can have
typical time constants on the order of several tens of ns (see
Appendix), fast enough to compete with the time scale of
radiation pressure effects on the order of $1/(2 \kappa)$ (approx.
13 ns in our experiment). The direction of the force depends on
the specific material properties of the expanding layers. In our
case and in contrast to previous experiments it assists the
cooling effect of radiation pressure present for positive
detuning.

The experimental data is consistent with radiation-pressure
cooling assisted by photothermal effects. Residual heating of the
cantilever due to absorption could not be observed (see
Appendix). Improvements of the Bragg mirror reflectivity will
further reduce and eventually eliminate photothermal
contributions to the cooling since it will allow to achieve a
higher finesse and to limit the optical absorption.   An
interesting analogy to understand this cooling mechanism can be
found in thermodynamics. If a system (the mirror), initially at
thermal equilibrium with a bath at ambient temperature (its
environment), is strongly coupled to another bath with a very low
temperature (the low-noise laser), its temperature will decrease
in order to bring it to equilibrium with both baths. The current
technical limitation for observing a lower temperature is the
stability of the detuned locking and the base temperature from
which the self-cooling starts. For example, with a cavity finesse
$F=6000$ and 1~mW of input optical power we expect a pure
radiation pressure cooling ratio of 1500 for a smaller mirror
oscillating at 1~MHz with an effective mass of 5~ng and $Q=10^5$.
Starting from 5~K, one should achieve cooling to 3~mK, below the
base temperature of a dilution fridge. We are confident that the
quantum ground state may be reachable with state-of-the-art optics
and microfabrication technique \cite{Pinard2005b}.

\section{Conclusion}
We have observed self-cooling of a micro-mirror sustained by
radiation pressure. The cooling of mechanical oscillators is a key
requirement for many open problems of modern physics ranging from
the performance of shot-noise limited position measurements
\cite{LaHaye2004} to the study of gravitational waves
\cite{Braginsky2002,LIGO} and dynamical multistability in
micro-optical systems \cite{Marquardt2005}. The possibility of
lowering the temperature of an oscillator to its quantum mechanical
ground state paves the way to the implementation of quantum state
engineering involving macroscopic systems
\cite{Braunstein2003,Bose1997,Mancini1994}, a closer study of the
boundary between classical and quantum physics
\cite{Bouwmeester2003} and, ultimately, the observation of
non-classical correlations between macroscopic objects
\cite{Pinard2005b}. In the long run it may also provide new means
for integrated quantum (mechanical) information processing.

\begin{acknowledgments}
We would like to thank C. Brukner, S. Gr\"oblacher, J. Kofler, T. Jennewein, M. S. Kim, and D. Vitali for discussion. We acknowledge financial support by the Austrian Science Fund (FWF), by the City of Vienna and by the Austrian NANO Initiative (MNA).
\end{acknowledgments}

\newpage
\appendix
\section{Appendix}

\subsection{Mechanical characterization of the vibrational mode}
\subsubsection{CANTILEVER CHARACTERISTICS}
The doubly clamped cantilever is a free standing Bragg mirror made
of several layers of $TiO_2$/$SiO_2$. The characteristics of the
Bragg mirror are summarized  in table \ref{materialProperties}.

\begin{table}[htb]
\caption{Properties of the Bragg mirror}
\begin{tabular}{|c|c|c|c|c|c|}
\hline
Material    & Density & Layer  &     Number  & Index of & Bulk thermal   \\
 & ($kg/m^3$) & Thickness  ($nm$) & of layers & refraction & diffusivity $cm^2/s$ \\
\hline
$SiO_2$ &   2200 &  183.45  & 8 & 1.45 &    0.086 \\
\hline
$TiO_2$ &   4200 &  107.26  & 9 & 2.48 &    0.031 \\
\hline
\end{tabular}
\label{materialProperties}
\end{table}

The dimensions of the structure are set by the laser-ablated pattern
to 490~$\mu$m x 110~$\mu$m, which corresponds to a total mass of
390~ng. However the ends of this structure are not totally undercut
and therefore not free to move. This means that the actual
cantilever is shorter and the total mass participating in the
oscillation is lower.

\subsubsection{MODE TOMOGRAPHY}

The mechanical mode has been spatially characterized by scanning the
surface of the cantilever and performing point-by-point measurements
of the mean square displacement at the position of the optical beam.
The measurement has been done at low power (input power  $\approx$
200~$\mu$W) and at zero detuning to avoid spurious
radiation-pressure effect. By performing a longitudinal scan
(approximately 40 points regularly spaced by 12~$\mu$m) along a line
close to the center of the bridge, we reconstructed the transverse
profile of the first three modes and compared it with theoretical
simulations (Fig. \ref{modes}):

\begin{figure}[htb]
\centerline{\includegraphics[width=0.5\textwidth]{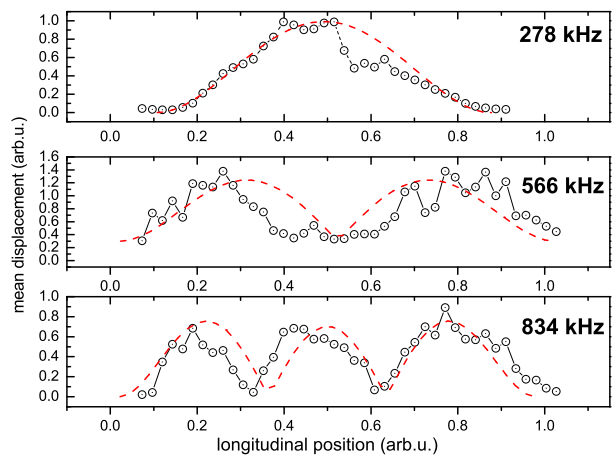}}
\caption{Mode shape of the first three modes of an ideal
doubly-clamped beam: Experimental points (dots) and comparison with
the theoretical shape for the first three longitudinal modes of a
doubly clamped bridge (red line).} \label{modes}
\end{figure}

From this analysis, we conclude that the oscillation along the
longitudinal direction of the mode at 280~kHz is a fundamental mode.
The frequency pattern is consistent with a tension-dominated
vibration since higher order mode frequencies are almost harmonics
of the fundamental one (i.e. close to the ratio 2:1 and 3:1
respectively).

\begin{figure}[htb]
\centerline{
\includegraphics[width=0.4\textwidth]{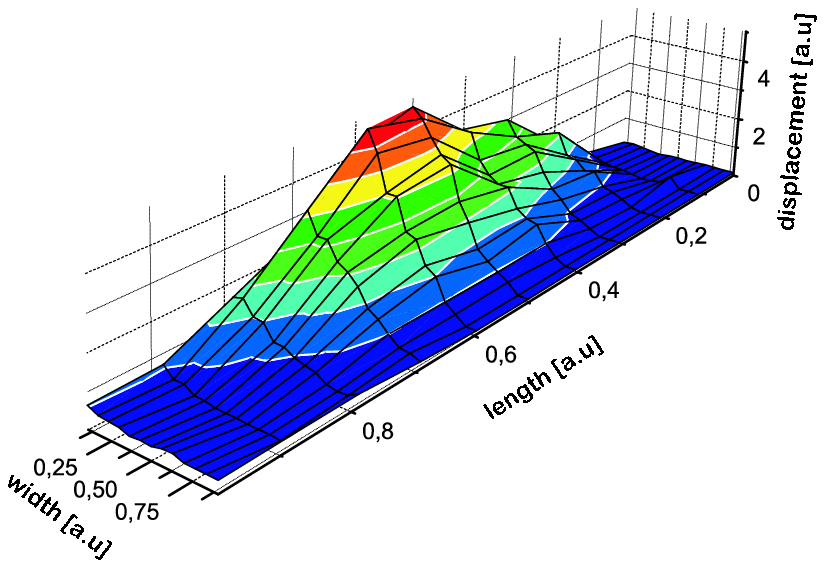}
\includegraphics[width=0.4\textwidth]{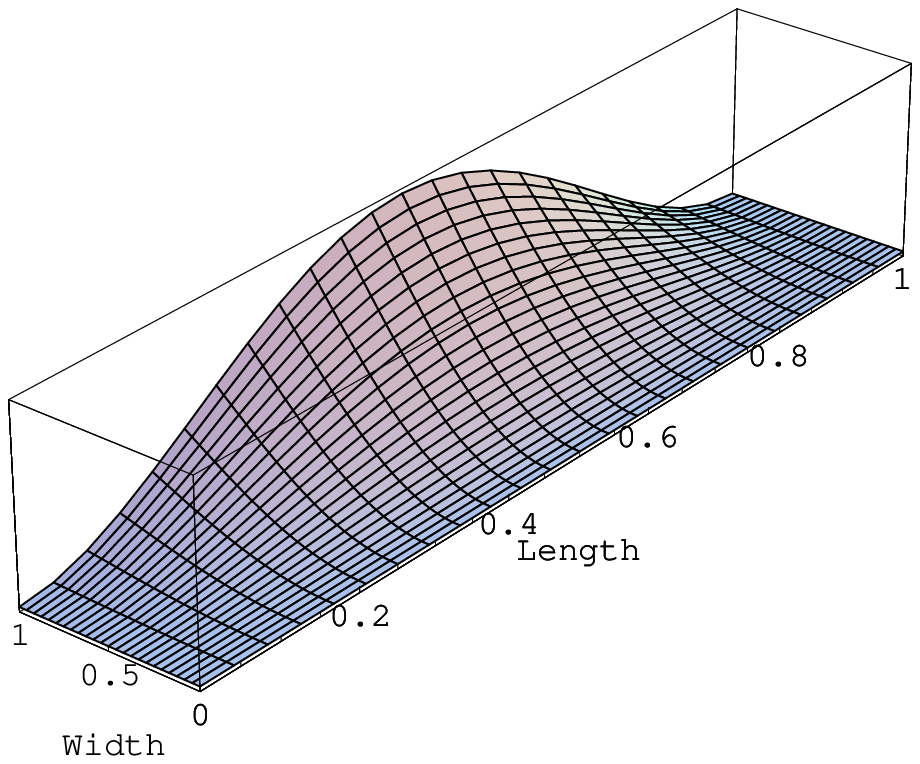}}
\caption{ 2D tomography of the mean displacement of the 280~kHz
vibrational mode (measurement (left) and simulation (right) )}
\label{tomography}
\end{figure}

In order to determine the effective mass relevant for
radiation-pressure effects (see below), we have performed a 2D
tomography (approximately 15 x 10 points, corresponding to a mesh of
roughly 10~$\mu$m x 50~$\mu$m) of the observed mechanical mode at
280kHz (Fig. \ref{tomography}). We see a clear decrease in the
amplitude of oscillation when moving laterally towards one side of
the cantilever. It is important to note that the coating is slightly
damaged by the laser ablation close to this side of the structure.
This means that approximately 30\% of the cantilever cannot be
addressed. From the mode shape we conclude that this part does not
participate in the vibration of the mode. This spatial behavior has
been modeled by considering a transverse behavior consistent with an
additional clamping on one side, as shown in Fig. \ref{tomography}.
The agreement in the shape of the mode is evident.

The cantilever does not behave as an ideal doubly-clamped cantilever
but as if approximately half of it is not resonating at 280~kHz,
resulting in an apparent clamping on one side.

\subsubsection{EFFECTIVE MASS}

It is not possible to simply consider the probed mechanical mode as
a point-like oscillator and to associate to it the total mass of the
object. One has to take into account the spatial point and extent
over which the probing is done, since the amplitude of the
vibrations is position-dependent. In other words, the mass which
enters in the mechanical susceptibility
$x[\Omega]=\chi[\Omega]F[\Omega]$, written as:
\begin{equation*}
\chi[\Omega]= \cfrac{1}{M_{eff}\left(\Omega_M^2 - \Omega^2 -i \cfrac{\Omega_M\Omega}{Q} \right)}
\end{equation*}
is not the full mass but an "effective" one, $M_\text{eff}$, which
has to be carefully evaluated. A full formal treatment taking this
effect into account has been developed and experimentally
demonstrated in \cite{Pinard1999,Briant2003}. We will explain it
briefly, in order to provide an independent evaluation of
$M_\text{eff}$ in our particular case.

For a point-dependent force $F(r,t)$, acting so to excite a
displacement mode of the cantilever normal to its surface, the total
work can be written as $W(t)=-<F(r,t),u(r,t)> $, where the bracket
stands for a spatial integration over the cantilever's surface. For
radiation pressure force from an optical beam with spatial profile
$v(r)$ and intensity $I(t)$, the force can be written as:
\begin{equation*}
F_{rad}=2\hbar k v^2(r)I(t)
\end{equation*}
where we have normalized the profile v as $<v^2>=1$ . The effective
susceptibility $\chi_\text{eff}$ involves then the effective mass
$M_\text{eff}$  reading
$M_\text{eff}=\rho_s\cfrac{<u^2>}{<u,v^2>^2}$ with $\rho_s$ the
surface density. This definition is independent of the normalization
of $u(r,t)$ . If one considers a point-like probing at position
$r_0$, the term $<u^2>/<u,v^2>^2$ reduces to $<u^2>/u^2(r_0)$, which
is the ratio between the mean square displacement at point $r_0$ and
the square displacement over the surface of the mode. In other
words, this formalizes the localization of the mode. This is even
clearer when one considers a higher-order mode of oscillation. In
this case the mode shape presents nodes (anti-nodes) of oscillation,
corresponding to points or regions where the amplitude is zero
(maximum). By probing one particular mode on its node, the amplitude
of the oscillation will be zero, corresponding to an infinite
effective mass. On the other hand, if one has to probe at an
antinode, the effective susceptibility to the probe is maximally
enhanced compared to the ideal point-like oscillator, so that
$M_\text{eff}$ can be strongly reduced with respect to the total
mass.

From the tomography of the mode (Fig. \ref{tomography}), we
obtain a lower bound (at the antinode) for the mass magnification
ratio $<u^2>/u^2(r_0)$ of 1/10. The estimation of this factor is
limited by the coarse-grained tomographic reconstruction of the
mode since it depends on the square of the mean displacement. The
estimation of the exact mass participating to the oscillations,
for the different reasons stated above, is difficult. However,
there is clear evidence that a potentially large fraction of the
bridge does not contribute to the total mass (reasonably
estimated to $\approx$50\% of the mass). Taking the full mass of
the bridge as an upper bound for the total mass and 1/10 as a
lower bound for the magnification ratio due to the localization
of the mode, we obtain in 39 ng a very conservative upper bound
for the effective mass, and a realistic evaluation for the
effective mass around 20 ng.

We have performed an independent evaluation of the effective
mass. Assuming room temperature (300 K), we can deduce via the
equipartition theorem from the thermal motion of the cantilever
the effective mass at the probing point. We send 0.1 mW of power
in the cavity and measure the noise power of the light in a
frequency window around 280 kHz. By knowing the cavity
characteristics and by carefully calibrating the electronic
detection scheme, one obtains from the measured intensity
modulation the (calibrated) mean square displacement of the
mirror and, via the equipartition theorem, the effective mass.
Our analysis yields a value of $22\pm 4$ ng at the antinode.

\subsection{Cavity characteristics}

The input cavity is a massive mirror (0.5" diameter, with radius
of curvature R=25~mm) with a measured reflectivity of
99.3$\pm$0.2\%. The reflectivity of the micro-mirror is evaluated
to be 99.7$\pm$0.2\% by measuring the cavity finesse outside of
the bridge. On the cantilever the finesse is reduced, probably
due to scattering and diffraction losses close to the edges of the
cantilever. The cavity length is chosen to be close to a
semi-concentric configuration (consisting of a length slightly
smaller than 25mm which characterizes semi-concentricity). In this
configuration, the size of the cavity waist is very sensitive to
minute changes of the cavity length and can be adjusted to be much
smaller than the bridge width, so as to reduce scattering and
edge-diffraction. By careful positioning on the cantilever, one
can obtain a finesse as large as 500, corresponding to a waist
diameter of approximately 20~$\mu m$ and to roughly  0.2\% extra
losses.

\subsection{Assessement of the photothermal contribution}
\subsubsection{Manifestation of Heating}
\begin{figure}[htb]
\centerline{\includegraphics[width=0.5\textwidth]{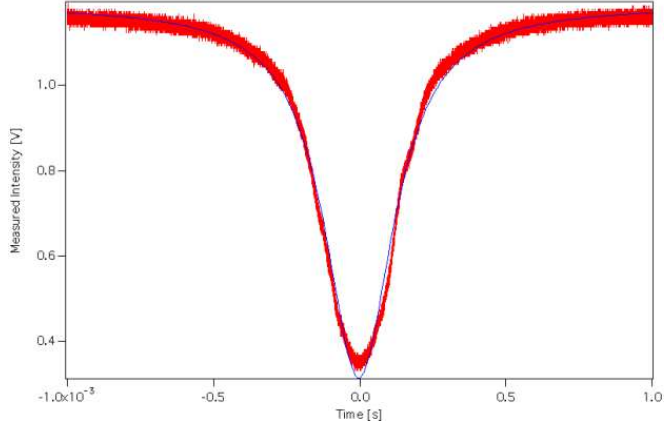}} \caption{
Experimental length scan of the cavity around a reflection peak
for an input power of  2mW (red), and Lorentzian fit (blue). ).
The cavity is slowly scanned in time. The FWHM of the lorentzian
correspond to  $\lambda/F \approx 2nm$} \label{peak}
\end{figure}

When scanning the length of the cavity over more than a free
spectral range (by sending a ramp voltage to the piezo electric
transducer on which the massive input mirror is placed), one can
record the intensity reflected from the cavity (see Fig. \ref{peak}
for the reflection peak corresponding to an input power of 2~mW) and
measure the finesse. This is given by the width of the reflection
peak divided by the free spectral range. From Fig. \ref{peak}, we
can see that there is no clear asymmetry in the peak shape, whose
presence would be a manifestation of strong heating effects.

\begin{figure}[htb]
\centerline{\includegraphics[width=0.5\textwidth]{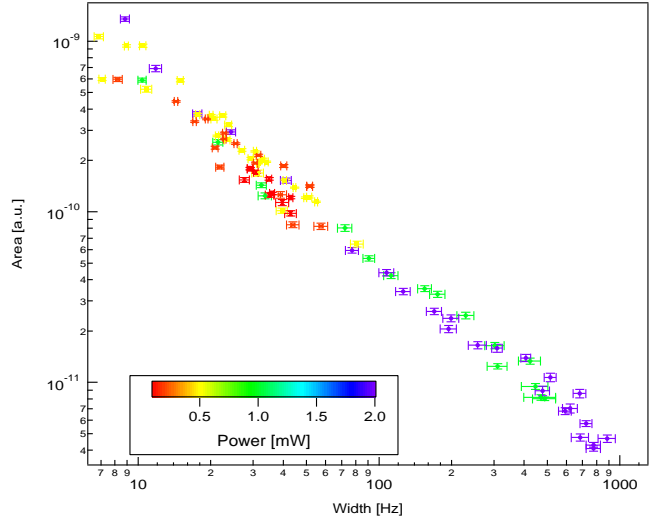}}
\caption{Area versus width for different input power (see text).}
\label{areaVsWidth}
\end{figure}

It is possible to evaluate the photo-absorption and the consequent
heating of the cantilever by comparing the temperature of the mode,
proportional to the area underneath a mechanical resonance peak, and
the corresponding width, for different input powers. The results are
summarized in Fig. \ref{areaVsWidth}.

By looking at the behavior corresponding to a given input power, we
see that width and temperature are inversely related, as expected
from the self-cooling mechanism. However, by comparing the curves
obtained at different input power, we can also gather information
about the heating of the cantilever. If no residual heating takes
place, all points will fall on the same line irrespective of the
laser power, since cooling is only given by the effective damping in
the system. Heating would be revealed by a shift of the data set
belonging to a certain laser power as compared to a data set at
lower laser power. On the contrary, the experimental curves in Fig.
\ref{areaVsWidth} are, within the error bars, superimposed  in a way
that the described stacking cannot be observed. This implies there
is no discernable heating of the cantilever within the range of
input powers we have considered. Furthermore,  Fig.
\ref{areaVsWidth} shows  that even if some residual heating is
present (which, within the error bar, is not the case), it does not
undermine the self-cooling performance in our experiment.

\subsubsection{Timescale of the photothermal effects}
The mechanical and thermal properties of our free standing layered
structures are not easy to assess. Mechanical and thermal properties
 of thin layered microstructures are not well known and can strongly
 differ from the those of bulk materials \cite{Lee1995}. It is therefore
  difficult to measure or calculate the exact timescale and strength of
  the photothermal effects in our experiment. The inferred strength of
   the photothermal contribution to the  cooling of the micro-mechanical
    oscillator suggests that the effect is at the same time strong and
     very fast (i.e. much faster than the period of oscillation of the
      mode, and in competition with the typical radiation pressure
      response time $1/\kappa$=13~ns). Usually, photothermal processes
       are typically relatively slow processes. However, we will give
       a simple and rough evaluation of the thermalisation time, showing
       that a timescale as fast as the radiation pressure response time,
       in our conditions, is easily achievable. We can consider that the
lateral dimension of the heat-affected zone on the micromirror is
roughly the size of the waist of the cavity (20~$\mu$m). The
evanescent wave on the Bragg mirror has a penetration depth which
is below 0.4 microns. This means that the heating due to
absorption mainly affects the first couple of layers in the
structure. Due to different
 absorption coefficients, these first layers heat up differently over
  their whole lateral dimension. As a mechanism for thermalisation,
  we can therefore restrict ourselves to a one-dimensional heat
  diffusion process between the two first layers of material that is causing
  the thermalization. The timescale of the fastest photothermal
   forces should be of the order of this thermalisation time,
   assuming an initial difference of temperature between the
    two first layers. Assuming that the usual relation for
    the heat diffusion length $l_T=\zeta\sqrt{D\tau}$ holds
     for such thin layers (where $l_T$ is the diffusion
     length after a time $\tau$ , $\zeta$  is a geometrical
      factor roughly in the order to of unity, and $D$ is the thermal
       diffusivity), and assuming the mean bulk values for $D$ (see table
       \ref{materialProperties}), one gets an estimated thermalisation
       time for the two layers of typically about
       $\tau=\left(\cfrac{1}{\zeta}\right)^2\left(\cfrac{L_1^2}{D_1}+\cfrac{L_2^2}{D_2}\right)\approx$ 4~ns(with $L_1$
       and $L_2$  the thickness of the  two layers
        and $D_1$ and $D_2$ the thermal diffusivities of the
         two materials). The square dependence of the thermalisation time on the
         dimension of the structure allows for a relaxation time range of a few nanoseconds,
          which determine the contribution of this processes in the cooling mechanism.

\section*{REFERENCES}

\bibliography{refs-cooling}

\end{document}